\documentclass[twocolumn,showpacs,preprintnumbers,amsmath,amssymb]{revtex4}

\usepackage{graphicx}
\usepackage{dcolumn}
\usepackage{bm}
\usepackage{amsmath}
\usepackage{amssymb}

\begin{document}

\preprint{APS/123-QED}

\title{Hierarchical organization in complex networks}

\author{Erzs\'ebet Ravasz}
 \affiliation{Department of Physics, 225 Nieuwland Science Hall,
 		 University of Notre Dame, Notre Dame, IN 46556, USA}
\author{Albert-L\'aszl\'o Barab\'asi}
\affiliation{Department of Physics, 225 Nieuwland Science Hall,
 		 University of Notre Dame, Notre Dame, IN 46556, USA}

\date{\today}

\begin{abstract}
Many real networks in nature and society share two generic properties: they are scale-free
and they display a high degree of clustering.
	 We show that these two features are the consequence of a hierarchical organization,
implying that small groups of nodes organize in a hierarchical manner into increasingly
large groups, while maintaining a scale-free topology.
	In hierarchical networks the degree of clustering characterizing the
different groups follows a strict scaling law, which
can be used to identify the presence of a hierarchical organization in real networks.
	We find that several real networks, such as the World Wide Web,
actor network, the Internet at the domain level and the semantic web
obey this scaling law, indicating that hierarchy is a fundamental
characteristic of many complex systems.
\end{abstract}

\pacs{89.75.-k, 89.20.Hh, 05.65.+b}
\maketitle

	In the last few years an array of discoveries have redefined our understanding of
complex networks (for reviews see \cite{reka_rev,dorog_rev}).
	The availability of detailed  maps, capturing the topology of
such diverse systems as the cell \cite{metab_sf,ecoli,protein_sf,protein_wagner},
the world wide web \cite{19deg}, or the sexual network  \cite{sexweb},
have offered scientists for the first time the chance to address in quantitative terms
the generic features of real networks.
	As a result, we learned that networks are far from being random,
but are governed by strict organizing principles, that generate	systematic and measurable
deviations from the topology predicted by the random graph theory of
Erd\H os and R\'enyi   \cite{erdos,bollobas}, the basic model used to describe complex webs
in the past four decades.

	Two  properties  of real networks have generated considerable attention.
	First, measurements indicate that most networks display a high degree of clustering.
	Defining the clustering coefficient for node $i$ with $k_i$ links as $C_i=2n_i/k_i(k_i-1)$, where $n_i$
is the number of links between the $k_i$ neighbors of $i$, empirical results indicate that $C_i$ averaged over
all nodes is significantly higher for most real networks than for a random network of similar size  \cite{SW,reka_rev,dorog_rev}.
	Furthermore,  the clustering coefficient of real networks is to a high
degree independent of the number of nodes in the network (see Fig. 9 in   \cite{reka_rev}).
	At the same time, many networks of scientific or technological interest, ranging from the
World Wide Web \cite{19deg} to biological networks \cite{metab_sf,ecoli,protein_sf,protein_wagner} have been found to be
scale-free  \cite{BA_model,SF_Phys_A}, which means that the probability that a randomly selected
node has $k$ links (i.e. degree $k$) follows $P(k) \sim k^{-\gamma}$, where $\gamma$
is the degree exponent.

	The scale-free property and clustering
are not exclusive: for a large number of real networks, including metabolic
networks  \cite{metab_sf,ecoli}, the
protein interaction network   \cite{protein_sf,protein_wagner},
the world wide web  \cite{19deg} and even some
social networks  \cite{coauthor_newman1,coauthor_newman2,coauthor}
the scale-free topology and high clustering coexist.
	Yet, most models proposed to describe the topology of complex networks
have difficulty capturing simultaneously
these two features.
	For example, the random network model  \cite{erdos,bollobas}
cannot account neither for the scale-free, nor for the clustered
nature of real networks, as it predicts an exponential degree distribution, and
the average clustering coefficient, $C(N)$, decreases as $N^{-1}$ with the number of nodes in the network.
	Scale-free networks, capturing the power law degree distribution,
predict a much larger clustering coefficient than a  random network.
	Indeed, numerical simulations indicate that
for one of the simplest models  \cite{BA_model,SF_Phys_A}
 the  average clustering coefficient depends on the system size as
$C(N)\sim N^{-0.75}$  \cite{reka_rev,dorog_rev},  significantly larger for large $N$
than the random network prediction $C(N)\sim N^{-1}$.
	Yet, this prediction still disagrees with the finding that for several real systems
 $C$ is independent of $N$ \cite{reka_rev}.

	Here we show that the fundamental discrepancy
between models and empirical measurements is rooted in a previously
disregarded, yet generic feature of many real networks: their hierarchical topology.
	Indeed, many networks  are fundamentally modular: one can easily identify groups
of nodes that are highly interconnected with each other, but have only a few or no links to nodes
outside of the group to which they belong to.
	In society such modules represent groups of friends or coworkers \cite{granovetter};
	in the WWW  denote communities with shared interests
 \cite{giles_web_clust,friends_on_web_Adamic};
in the actor network they characterize specific genres or simply individual movies.
	Some groups are small and tightly linked, others are larger and somewhat
less interconnected.
	This clearly identifiable modular organization
is  at the origin of the high clustering coefficient seen in many real networks.
	Yet, models reproducing the scale-free property of real networks  \cite{reka_rev,dorog_rev} distinguish nodes
based only on their degree, and are blind to node characteristics that could lead to
a modular topology.

	In order to bring modularity,
the high degree of clustering and the scale-free topology under a single roof,
we need to assume that modules combine into each other in a
hierarchical manner, generating what we call a \emph{hierarchical network}.
	The presence of a hierarchy and the scale-free property impose strict restrictions on the number
and the degree of cohesiveness of the different groups present in a network, which can be captured
in a quantitative manner using a scaling law, describing the dependence of the clustering
coefficient on the node degree.
	We use this scaling law to identify the presence of a hierarchical architecture in several real networks,
and the absence of such hierarchy in  geographically organized webs.

\section{ Hierarchical Network Model}

\begin{figure}[h]
\centering
\begin{minipage}[c]{0.5\textwidth} 
\centering
    \begin{minipage}[c]{0.27\textwidth}
	\centering

\vspace{0.5cm}
	\includegraphics[width=0.5\textwidth]{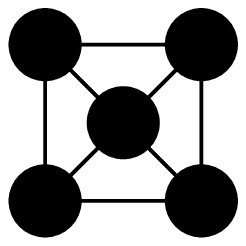}

	\footnotesize
	{\bf (a)} n=0, N=5

\vspace{1.3cm}
	\includegraphics[width=\textwidth]{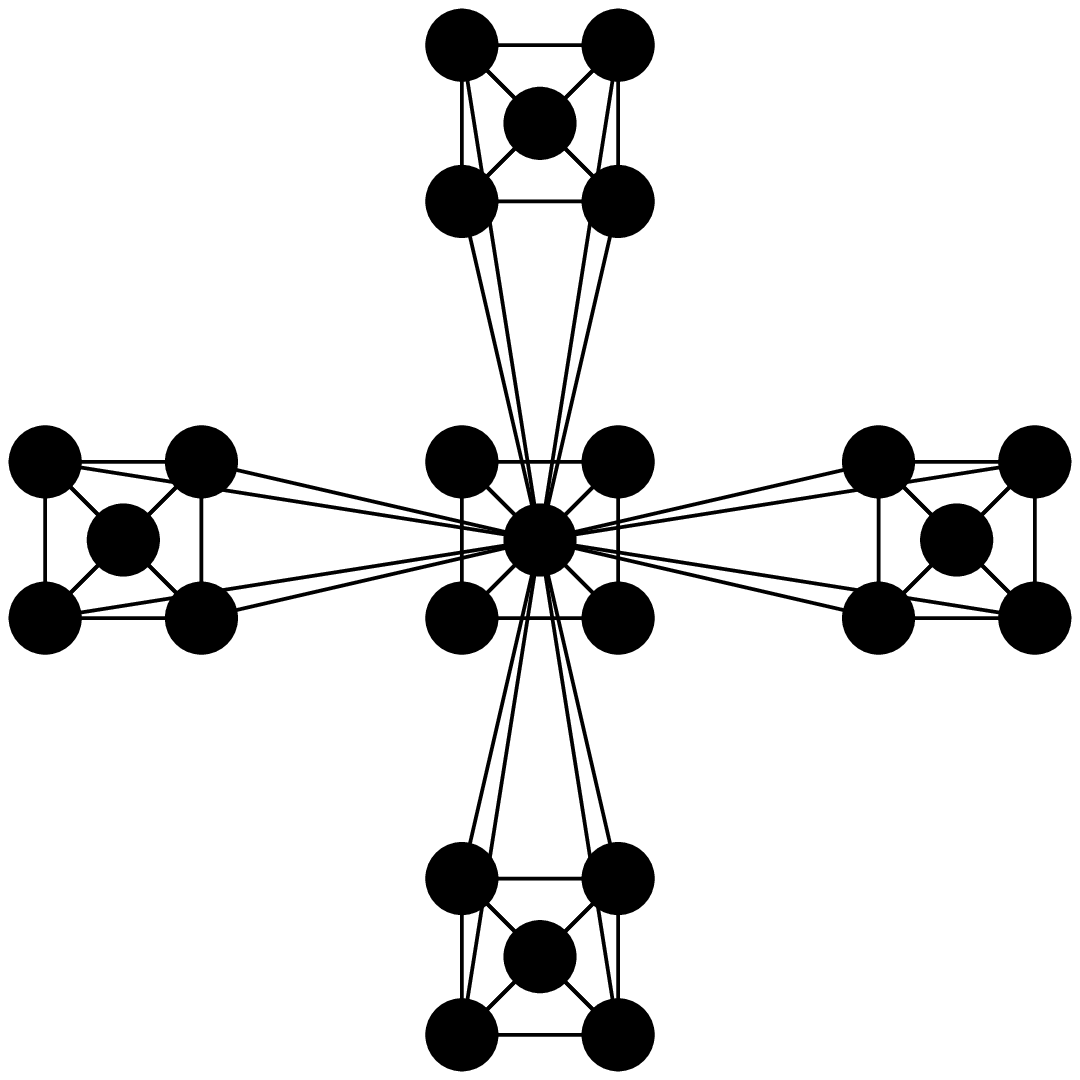}

	\footnotesize
	 {\bf (b)} n=1, N=25
     \end{minipage}
\hspace{0.005\textwidth}
     \begin{minipage}[c]{0.66\textwidth}
	\centering
	\includegraphics*[width=\textwidth,bb=200 200 1300 1300]{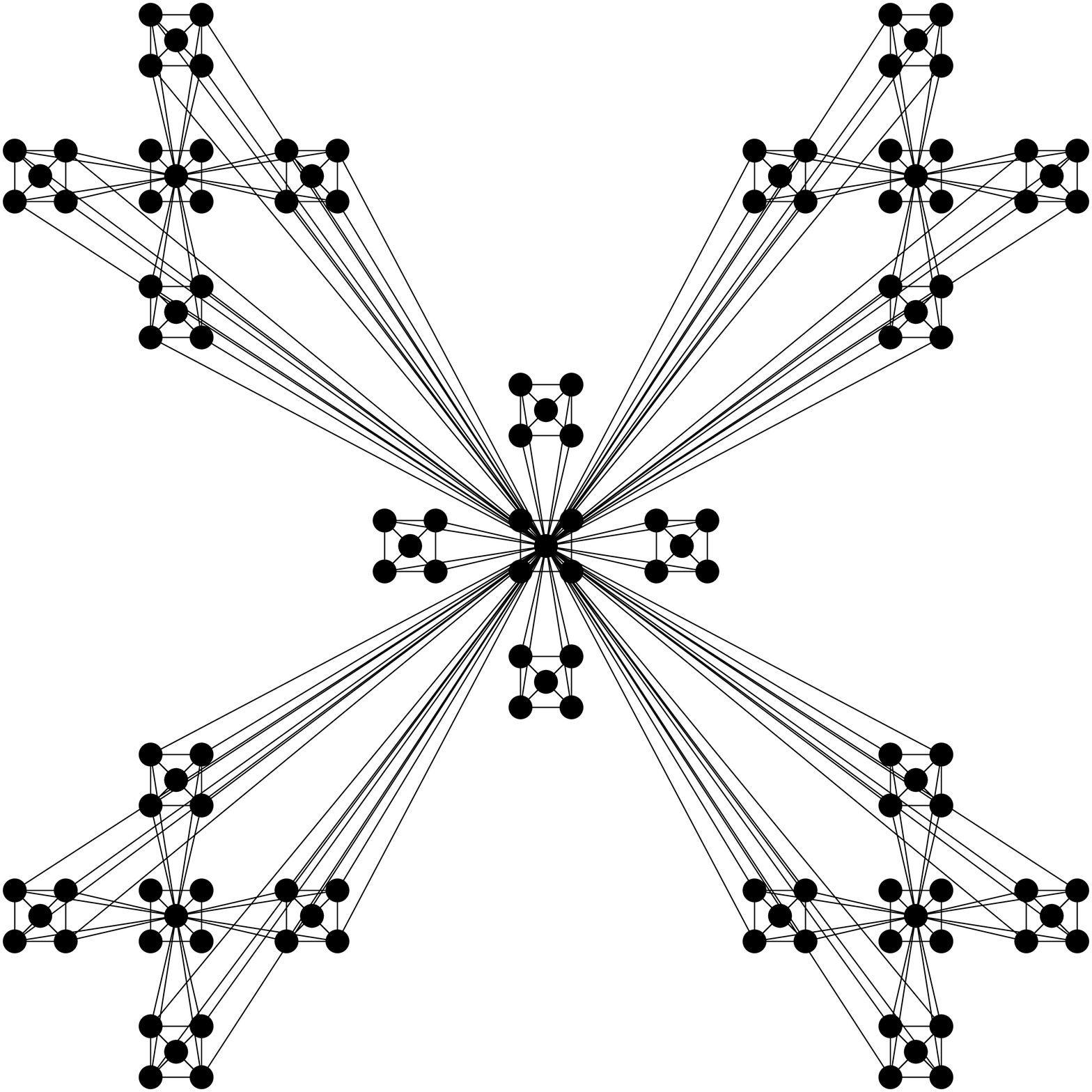}

	\footnotesize
	{\bf (c)} n=2, N=125
     \end{minipage}
\end{minipage}
\caption{
The iterative construction leading to a hierarchical network.
		Starting from a fully connected cluster
of five nodes shown in {\bf (a)} (note that the diagonal nodes are also connected  -- links not visible),
we create four identical replicas, connecting the peripheral nodes of each
cluster to the central node of the original cluster, obtaining a network of $ N= 25$ nodes {\bf (b)}.
In the next step we create four replicas of the obtained cluster, and connect the peripheral nodes again,
as shown in {\bf (c)}, to the central node of the original module, obtaining a $N = 125$ node network.
	This process can be continued indefinitely.
	}
\label{fig:det_model}
\end{figure}

	We start by constructing a hierarchical network model, that combines the scale-free
property with a high degree of clustering.
	Our starting point is a small cluster of five densely linked nodes
(Fig.$\,$\ref{fig:det_model}a).
	Next we generate four
replicas of this hypothetical module and connect the four external nodes of the
replicated clusters to the central node of the old cluster, obtaining a large  25-node
module (Fig.$\,$\ref{fig:det_model}b).
	Subsequently, we again generate  four
replicas of this 25-node module, and connect the 16 peripheral nodes to
the central node of the old module (Fig.$\,$\ref{fig:det_model}c), obtaining a new module of 125 nodes.
	These replication and connection steps can be repeated indefinitely, in each step increasing the number
of nodes in the system by a factor five.

	Precursors to the model described in Fig.$\,$\ref{fig:det_model} have been proposed in
Ref.   \cite{det_hier} and extended and discussed in Ref.  \cite{DGM,Kahng_Fractal_SF}
as a method of generating deterministic scale-free networks.
	Yet, it was believed that aside from their deterministic structure,
their statistical properties are equivalent with the stochastic models
that are often used to generate scale-free networks.
	In the following we argue that  such hierarchical
construction generates an architecture that is significantly different
from the networks generated by traditional scale-free models.
	Most important, we show that the new feature of the model,
its hierarchical character, are shared by a significant number of real networks.

\begin{figure*}[t]
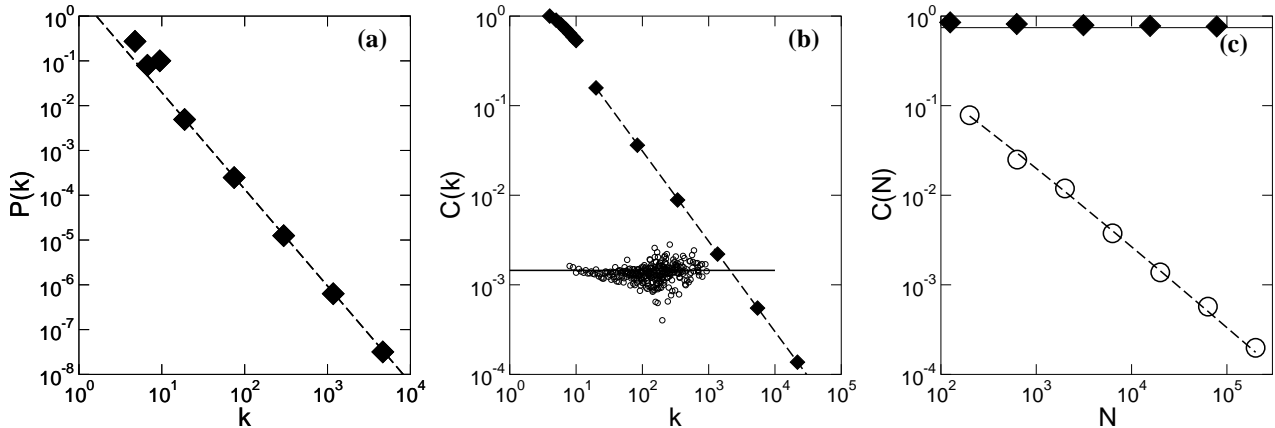

\centering
\begin{minipage}[c]{0.98\textwidth}
    \begin{minipage}[c]{0.32\textwidth}
	\centering
	\includegraphics*[width=\textwidth, bb = 195 0 700 550]{2a.eps}
     \end{minipage}
     \begin{minipage}[c]{0.32\textwidth}
	\centering
	\includegraphics*[width=\textwidth,bb = 195 0 700 550]{2b.eps}
     \end{minipage}
     \begin{minipage}[c]{0.32\textwidth}
	\centering
	\includegraphics*[width=\textwidth,bb = 195 0 700 550]{2c.eps}
     \end{minipage}
\end{minipage}

\caption{
	 Scaling properties of the hierarchical model shown in
			Fig.$\,$\ref{fig:det_model} ($N=5^7$).
	 {\bf (a)} The numerically determined degree distribution.
	 	The assymptotic scaling, with slope $\gamma=1+\ln5/\ln4$, is shown as a dashed line.
	 {\bf (b)} The $C(k)$ curve for the model, demonstrating that it follows Eq. (\ref{dgm_scale}).
	 	The open circles show $C(k)$ for a scale-free model  \cite{BA_model} of the same size,
		illustrating that it does not have a hierarchical architecture.
	 {\bf (c)} The dependence of the clustering coefficient, $C$, on the size of the network $N$.
	 	While for the hierarchical model $C$ is independent of $N$ ($\blacklozenge$),
		for the scale-free model $C(N)$ decreases rapidly ({\tiny $\bigcirc$}).
	}
\label{fig:det_model_graph}
\end{figure*}

	First we note that  the hierarchical network model seamlessly integrates a
scale-free topology with an inherent modular structure.
	Indeed, the generated network has a power law
degree distribution with degree exponent
$\gamma = 1 + \ln 5/\ln 4 = 2.161$ (Fig.$\,$\ref{fig:det_model_graph}a).
	Furthermore, numerical simulations indicate that the clustering coefficient, $C \simeq 0.743$, is
independent of the size of the network (Fig.$\,$\ref{fig:det_model_graph}c).
	Therefore, the high degree of clustering and the scale-free property are simultaneously
present in this network.

	The most important feature of the network model of Fig.$\,$\ref{fig:det_model},
not shared by either the scale-free  \cite{BA_model,SF_Phys_A}
or random network models  \cite{erdos,bollobas}, is its hierarchical architecture.
	The network is made of numerous small, highly
integrated five node modules (Fig.$\,$\ref{fig:det_model}a), which are assembled into larger
25-node modules (Fig.$\,$\ref{fig:det_model}b).
	These 25-node modules are less integrated but each of them
is clearly separated from the other 25-node
 modules when we combine them into the even larger
 125-node modules (Fig.$\,$\ref{fig:det_model}c).
 	These 125-node modules are even less cohesive,
but again will appear separable from their replicas if the network expands further.

	This intrinsic hierarchy can be characterized in a quantitative manner using the recent
finding of Dorogovtsev, Goltsev and Mendes  \cite{DGM} that in deterministic scale-free networks the
clustering coefficient of a node with $k$ links follows the scaling law
\begin{equation}\label{dgm_scale}
C(k) \sim k^{-1}.
\end{equation}

	We argue that this scaling law quantifies the coexistence of a hierarchy of
nodes with different degrees of clustering,
 and applies to the model of Fig.$\,$\ref{fig:det_model}a-c as well.
	Indeed, the nodes at the center of the numerous 5-node
modules have a clustering coefficient $C=1$.
Those at the center of a 25-node
module have $k = 20$ and $C = 3/19$,
while those at the center of the 125-node
modules have $k = 84$ and $C = 3/83$,
indicating that the higher a node's degree the smaller is its
clustering coefficient, asymptotically following the $1/k$ law (Fig.$\,$\ref{fig:det_model_graph}b).
	In contrast, for the scale-free model proposed in Ref.~ \cite{BA_model} the clustering coefficient
is independent of $k$, i.e. the scaling law (\ref{dgm_scale}) does not apply
(Fig.$\,$\ref{fig:det_model_graph}b).
	The same is true for the random  \cite{erdos, bollobas} or the various
small world models   \cite{SW,SW2},
for which the clustering coefficient is independent of the nodes' degree.

	Therefore, the discrete model of Fig.$\,$\ref{fig:det_model}
combines within a single framework the two key properties of real networks: their scale-free
topology and  high modularity, which results in  a system-size independent clustering coefficient.
	Yet, the hierarchical modularity of the model results in the scaling law (\ref{dgm_scale}), which
is not shared by the traditional network models.
	The question is, could hierarchical modularity, as captured by this model, characterize
real networks as well?

\section{Hierarchical Organization in Real Networks}

	To investigate if such hierarchical organization is present in real networks we measured the $C(k)$
function for several networks for which large topological maps are available.
	Next we discuss each of these systems separately.

\begin{figure}[h]
\centering
\begin{minipage}[c]{0.5\textwidth} 
\centering
    \begin{minipage}[c]{0.445\textwidth}
	\centering
	\includegraphics[width=\textwidth]{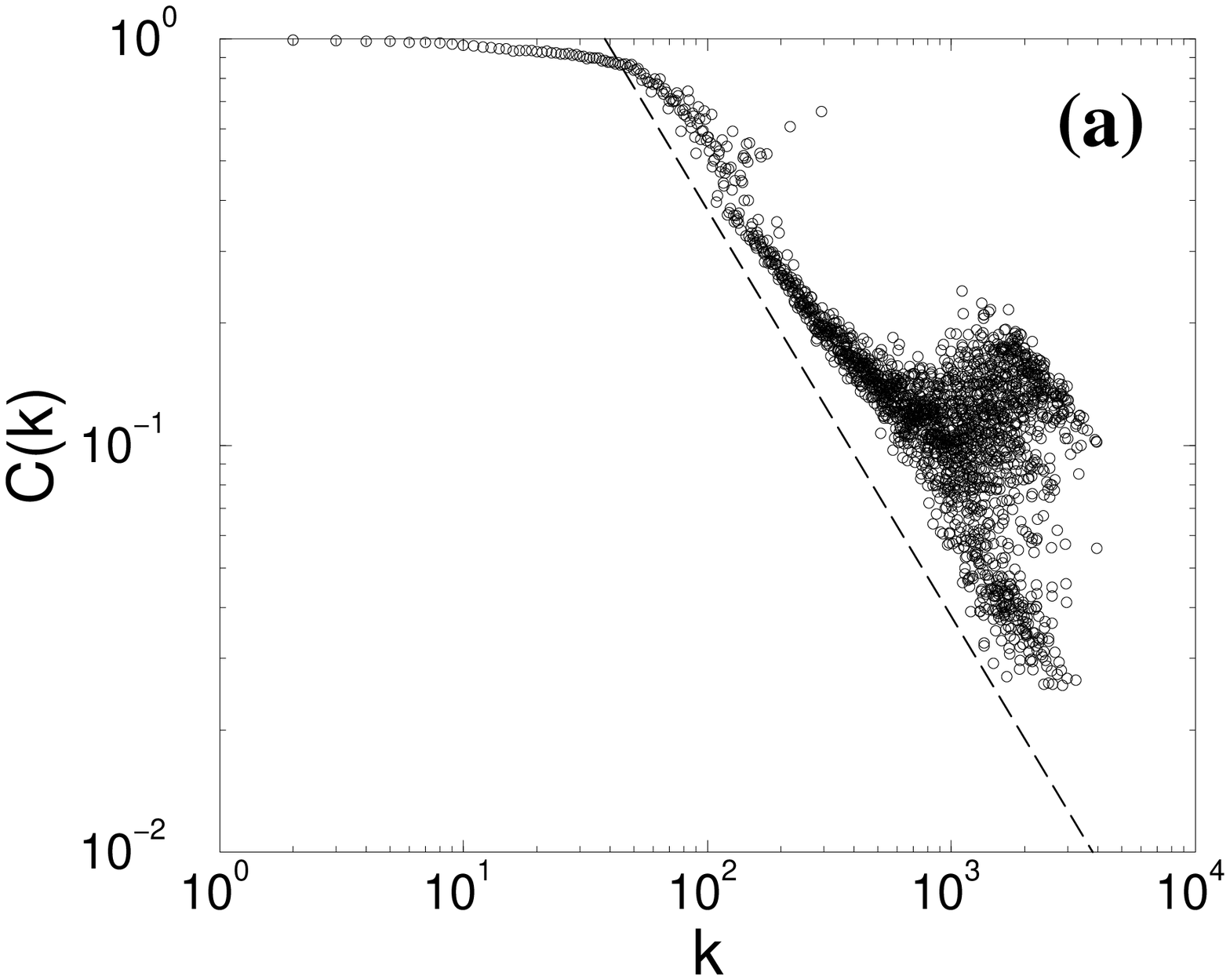}
     \end{minipage}
     \begin{minipage}[c]{0.445\textwidth}
	\centering
	\includegraphics[width=\textwidth]{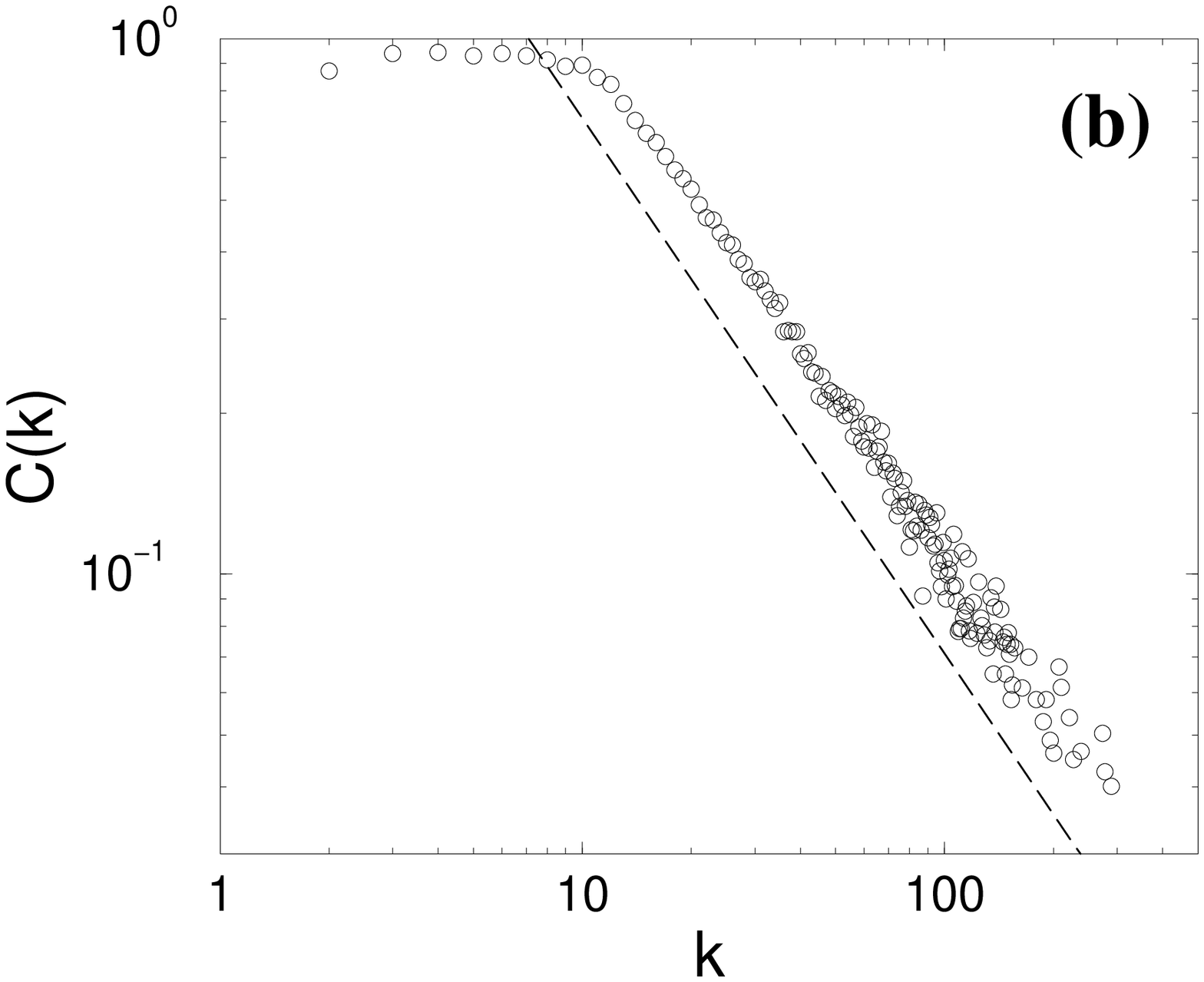}
     \end{minipage}
%
\centering
    \begin{minipage}[c]{0.445\textwidth}
	\centering
	\includegraphics[width=\textwidth]{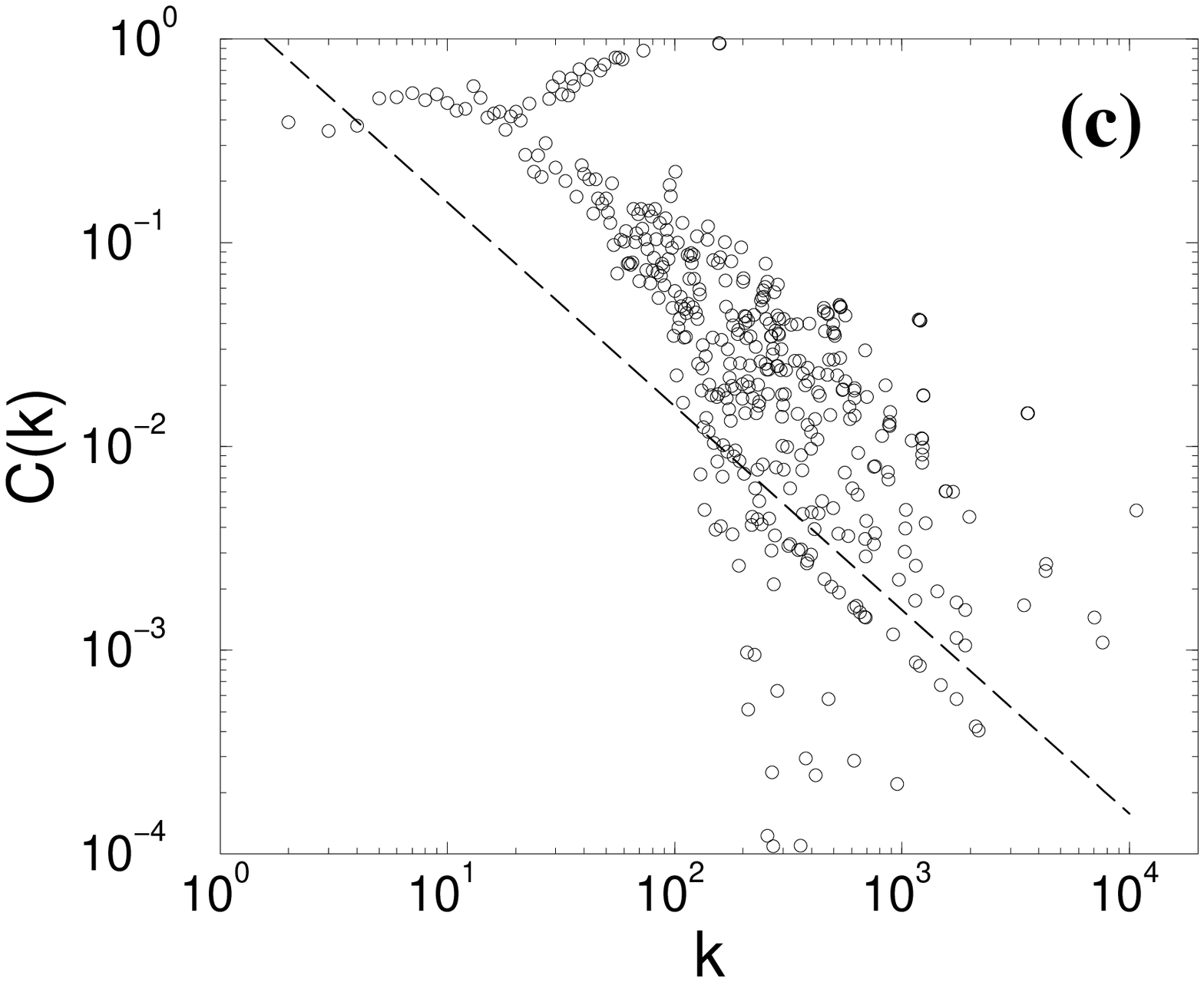}
    \end{minipage}
     \begin{minipage}[c]{0.445\textwidth}
	\centering
	\includegraphics[width=\textwidth]{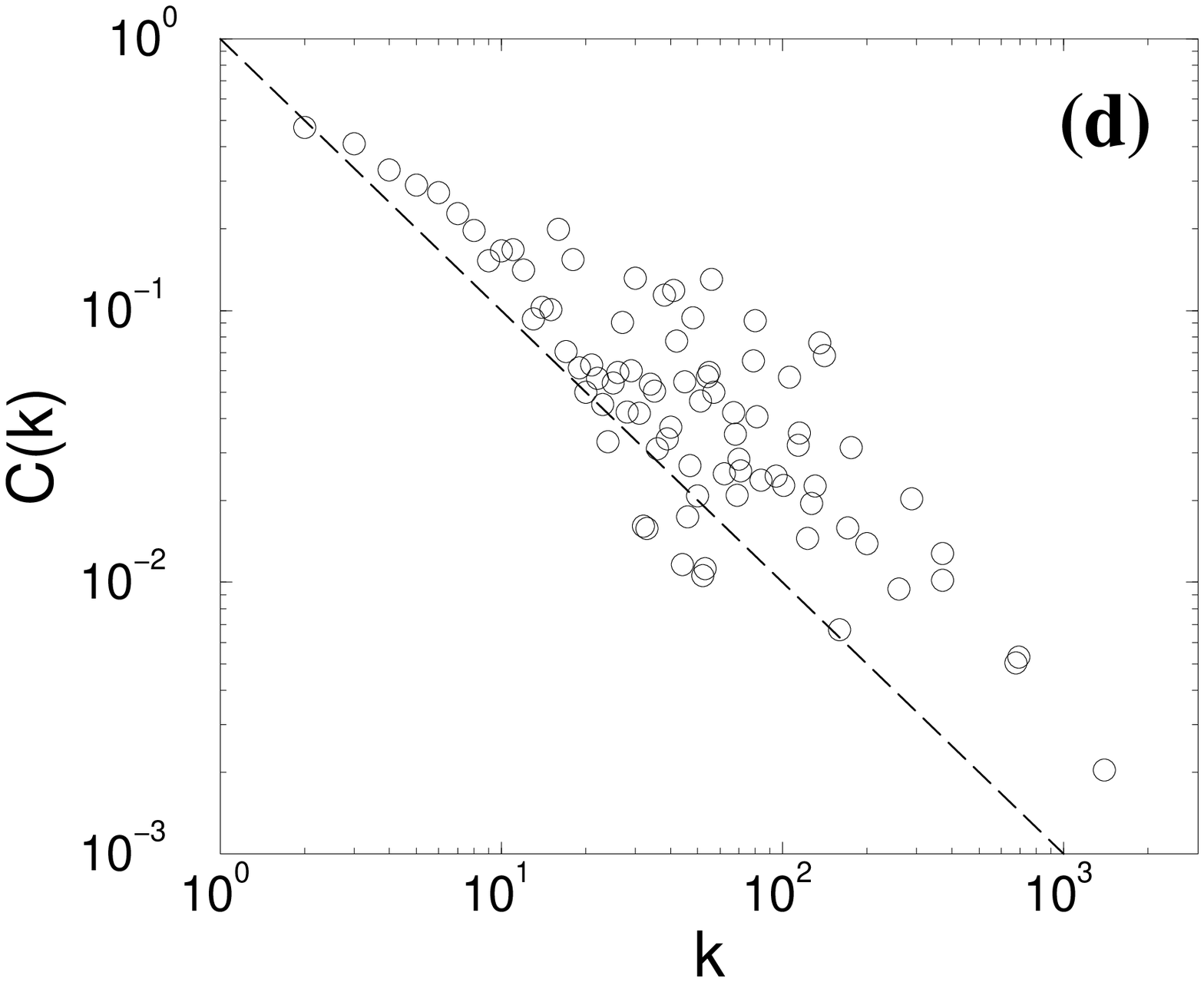}
     \end{minipage}
\end{minipage}

\caption{
	The scaling of $C(k)$ with $k$ for four large networks:
	{\bf (a)} Actor network, two actors being connected if they acted in the same movie according to the
		 {\texttt{www.IMDB.com}} database.
	{\bf (b)} The semantic web, connecting two English words if they are listed as synonyms in the
		Merriam Webster dictionary  \cite{syn}.
	{\bf (c)} The World Wide Web, based on the data collected in Ref.  \cite{19deg}.
	{\bf (d)} Internet at the Autonomous System level, each node representing a domain,
		connected if there is a communication link between them.
		  The dashed line in each figure has
		 slope $-1$, following Eq. (\ref{dgm_scale}).
	}
\label{fig:h_net}
\end{figure}

	\emph{Actor Network:} Starting from the \texttt{www.IMDB.com} database, we
connect any two actors in Hollywood if they acted in
the same movie, obtaining a network of 392,340 nodes and
15,345,957 links.
	Earlier studies indicate that this network is scale-free with an exponential cutoff in $P(k)$
for high $k$   \cite{BA_model,ext_BA_model,amaral_actor}.
	As Fig.$\,$\ref{fig:h_net}a indicates, we find that $C(k)$ scales  as $k^{-1}$,
indicating that the network has a  hierarchical topology.
	Indeed, the majority of actors with a few links (small $k$)
 appear only in one movie.
 	Each such actor has a clustering coefficient equal to one, as
all actors the actor has links to are part of the same cast, and are therefore connected to each other.
	The high $k$ nodes include many actors that
acted in several movies, and thus their neighbors are not necessarily linked
to each other, resulting in a smaller $C(k)$.
	At high $k$ the $C(k)$ curve splits into two branches, one of which continues
to follow Eq. (\ref{dgm_scale}), while the other saturates.
	One explanation of this split is the decreasing amount of datapoints available in this region.
	Indeed, in the high $k$ region the number of nodes having the same $k$ is rather small.
	If one of these nodes corresponds to an actor
that played only in a few movies with hundreds in the cast, it will have both high $k$
and high $C$, considerably increasing the average value of $C(k)$.
	The $k$ values for which such a high $C$ nodes are absent continue
to follow the $k^{-1}$ curve, resulting in jumps between the high and small $C$ values for large $k$.
	For small $k$ these anomalies are averaged out.

	\emph{Language network:} Recently a series of empirical
results have shown that the language, viewed as a network of words, has a scale-free topology
 \cite{words,syn,word_lex,dorog_word}.
	Here we study the network generated
connecting two words to each other if they appear as synonyms in the Merriam Webster dictionary  \cite{syn}.
	The obtained semantic web has 182,853 nodes and 317,658 links and
it is scale-free  with degree exponent $\gamma=3.25$.
	The $C(k)$ curve for this language network is shown in Fig.$\,$\ref{fig:h_net}b,
indicating that it follows (\ref{dgm_scale}), suggesting that the language has a hierarchical organization.

	\emph{World Wide Web:} On the WWW two documents are connected to each other
if there is an URL pointing from one document to the other one.
	The sample we study, obtained by mapping out the \texttt{www.nd.edu} domain   \cite{19deg},
has 325,729 nodes and 1,497,135 links, and it is scale-free with degree exponents
$\gamma_{\text{out}}=2.45$ and $\gamma_{\text{in}}=2.1$, characterising the out
and in-degree distribution, respectively.
	To measure the $C(k)$ curve we made the network undirected.
	While the obtained $C(k)$, shown in Fig.$\,$\ref{fig:h_net}c, does not follow as closely the scaling
law (\ref{dgm_scale}) as observed in the previous two examples, there is clear evidence that $C(k)$
decreases rapidly with $k$,
supporting the coexistence of many highly interconnected small nodes
with a few larger nodes, which have a much lower clustering coefficient.

	Indeed, the Web is full of groups of
documents that all link to each other.
	For example,  \texttt{www.nd.edu/$^\sim$networks}, our network research dedicated site,
has a  high clustering coefficient, as the
documents it links to have links to each other.
	The site is one of the several network-oriented sites, some of which point to each other.
	Therefore, the network research community  still forms a relatively cohesive group, albeit
less interconnected than the \texttt{www.nd.edu/$^\sim$networks} site, thus having a smaller $C$.
	This network community is nested into the much larger
community of documents devoted to  statistical mechanics, that
has an even smaller clustering coefficient.
	Therefore, the $k$-dependent $C(k)$ reflects the hierarchical nesting of the different
interest groups present on the Web.
	Note that $C(k)\sim k^{-1}$ for the WWW was observed and briefly noted in Ref.   \cite{wwwck}.

	\emph{Internet at the AS level:} The Internet is often studied at two different levels
of resolution.
	At the router level we have a network of routers connected by various
physical communication links.
	At the interdomain or autonomous system (AS) level each administrative domain,
composed of potentially hundreds of routers, is represented by a single node.
	Two domains are connected if there is at least one router that connects them.
	Both the router and the domain level topology have been found to be scale-free  \cite{faloutsos}.
 	As Fig.$\,$\ref{fig:h_net}d shows, we find that at the domain level  the Internet,
consisting of 65,520 nodes and 24,412 links   \cite{AS_data}, has a hierarchical topology
as $C(k)$ is well approximated with (\ref{dgm_scale}).
	The scaling of the clustering coefficient with $k$ for the Internet was earlier
noted by Vazquez, Pastor-Satorras and Vespignani (VPSV)
\cite{vespin_internet_1,vespin_internet_CK_2}, who observed $C(k)\sim k^{-0.75}$.
	VPSV interpreted this finding, together with the observation that the
average nearest-neighbor connectivity also follows a power-law with the node's
degree, as a natural consequence of the \emph{stub} and \emph{transit}
domains, that partition the network in a hierarchical fashion into
international connections, national backbones, regional networks
and local area networks.

	Our measurements indicate, however, that some
real networks lack a hierarchical architecture, and do not
obey the scaling law (\ref{dgm_scale}).
	In particular, we find that the power grid and the router level Internet topology
have a $k$ independent $C(k)$.

\begin{figure}[b] 
\begin{minipage}[c]{0.48\textwidth} 
\centering
    \begin{minipage}[c]{0.49\textwidth}
	\centering
	\includegraphics[width=\textwidth]{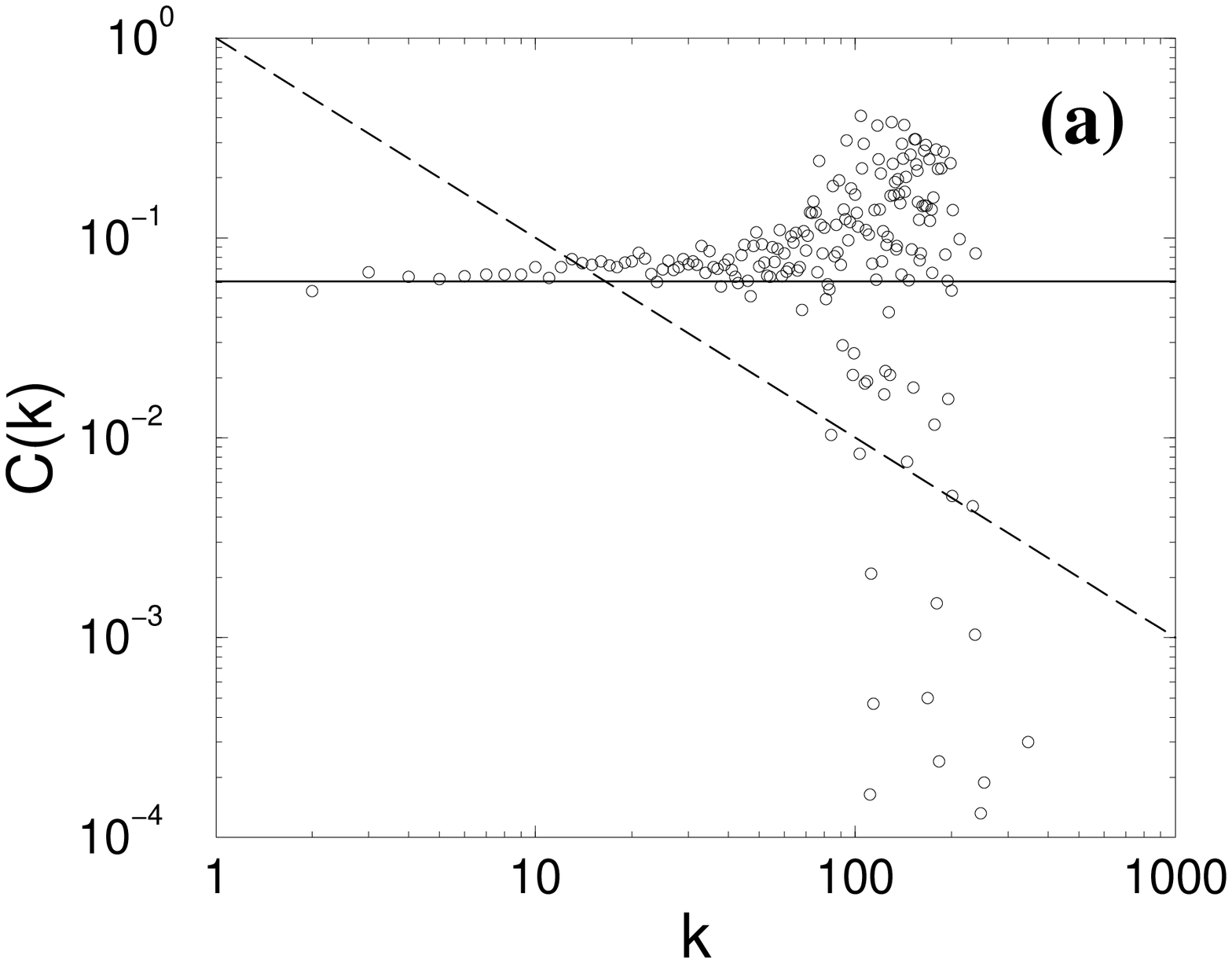}
%
     \end{minipage}
     \begin{minipage}[c]{0.49\textwidth}
	\centering
	\includegraphics[width=\textwidth]{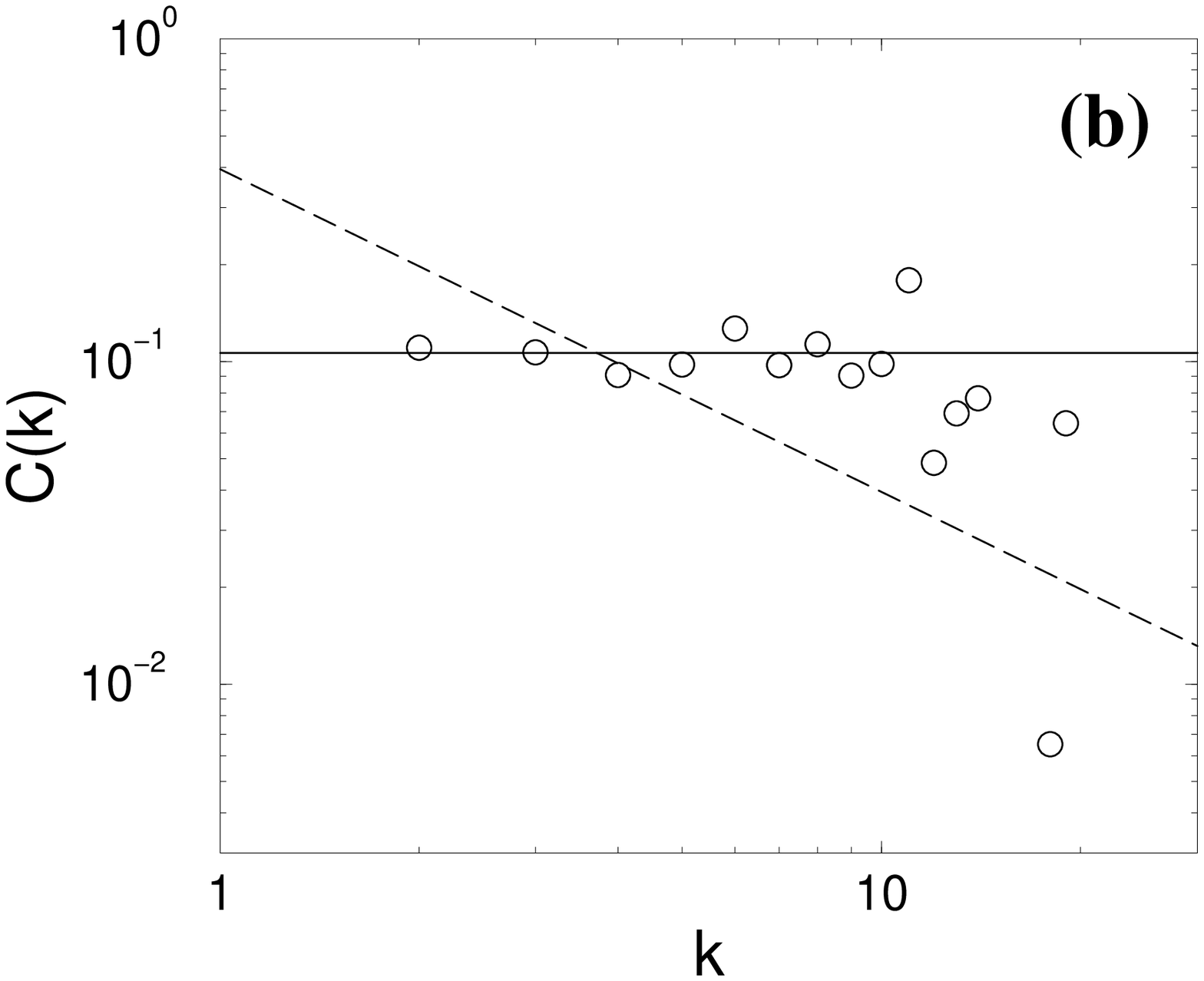}
%
     \end{minipage}
\end{minipage}
\caption{
 	The scaling of $C(k)$ for two large, non-hierarchical networks:
	{\bf (a)} Internet at router level  \cite{router_data}.
	{\bf (b)} The power grid of Western United States.
		 The dashed line in each figure has
		 slope $-1$, while the solid line
		 corresponds to the average clustering coefficient.
	}
\label{fig:hom_net}
\end{figure}

	\emph{Internet at the router level:} The router level Internet
has 260,657 nodes connected by 1,338,100 links  \cite{router_data}.
	Measurements indicate that the network is scale-free  \cite{faloutsos,int_model}
with degree exponent $\gamma=2.23$.
	Yet, the $C(k)$ curve (Fig.$\,$\ref{fig:hom_net}a), apart from some fluctuations,
is largely independent of $k$, in strong contrast with the $C(k)$
observed for the Internet's domain level topology (Fig.$\,$\ref{fig:h_net}d),
and in agreement with the results of VPSV
\cite{vespin_internet_1,vespin_internet_CK_2}, who also
note the absence of a hierarchy in router level maps.

	\emph{Power Grid:} The nodes of the power grid are generators, transformers and substations
and the links are high voltage transmission lines.
	The network studied by us
represents the map of the Western United States, and has 4,941 nodes and 13,188 links  \cite{SW}.
	The results again indicate that apart from fluctuations, $C(k)$
is independent of $k$.

	It is quite remarkable that these two networks share a common feature:
a geographic organization.
	The routers of the Internet and the nodes of the power grid have a well defined spatial
location, and the link between them represent physical links.
	In contrast, for the examples discussed in Fig.$\,$\ref{fig:h_net} the physical
location of the nodes was either undefined or irrelevant, and the length of the link was
not of major importance.
	For the router level Internet and the power grid the further
are two nodes from each other, the more expensive it is to connect them  \cite{int_model}.
	Therefore, in both systems the links are driven by cost considerations,
generating a distance driven structure, apparently excluding the emergence of
a hierarchical topology.
	In contrast, the domain level Internet is less distance driven, as many
domains, such as the AT\&T domain, span the whole United States.

   In summary, we offered evidence that for four large
 networks $C(k)$ is well approximated by $C(k) \sim k^{-1}$,
in contrast to
the $k$-independent $C(k)$ predicted by both the scale-free and random networks.
	In addition, there is evidence for similar scaling in the
metabolism \cite{Erzso_SCI_Ecoli} and protein interaction networks \cite{Soon-Hyung_protein}.
	This indicates that these networks have an inherently hierarchical organization.
	In contrast, hierarchy is absent in networks with strong geographical
constraints, as the limitation on the link length strongly constraints the network topology.

\section{Stochastic Model and Universality}

	The hierarchical model described in Fig.$\,$\ref{fig:det_model} predicts $C(k)\sim k^{-1}$,
which offers a rather good fit to three of the four $C(k)$ curves shown in Fig.$\,$\ref{fig:h_net}.
 	The question is, is this scaling law (\ref{dgm_scale}) universal, valid for all hierarchical networks, or
could different scaling exponent characterize the scaling of $C(k)$?
Defining the hierarchical exponent, $\beta$, as
\begin{equation}\label{beta}
	C(k) \sim k^{-\beta},
\end{equation}
is $\beta=1$ a universal exponent, or it's value can be
changed together with $\gamma$?
	In the following we demonstrate that the hierarchical exponent $\beta$ can be
tuned as we tune some of the network parameters.
	For this we propose a stochastic version of the model described in Fig.$\,$\ref{fig:det_model}.

	We start again with a small core of five nodes
all connected to each other (Fig.$\,$\ref{fig:det_model}a) and in step one ($n=1$)
we make four copies of the five node module.
	Next, we randomly pick a $p$ fraction of the newly added nodes
and connect each of them independently to the nodes belonging to the central module.
	We use preferential attachment  \cite{BA_model,SF_Phys_A} to decide to
which central node the selected nodes link to.
	That is, we assume that the probability that a selected node will connect to a node $i$
of the central module
is $k_i/\sum_j k_j$, where $k_i$ is the degree of node $i$ and the sum goes over all nodes of the central module.
	In the second step ($n=2$) we again create four identical copies of the 25-node
structure obtained thus far, but
we connect only a $p^2$ fraction of the newly added nodes to the central module.
 	Subsequently, in each iteration $n$ the central module of size $5^n$ is replicated four times, and in each
new module a $p^n$ fraction will connect to the current
central module, requiring the addition of $(5p)^n$ new links.

\begin{figure}[h]
\begin{minipage}[c]{0.5\textwidth} 
\centering
    \begin{minipage}[c]{0.49\textwidth}
	\centering
	\includegraphics[width=\textwidth]{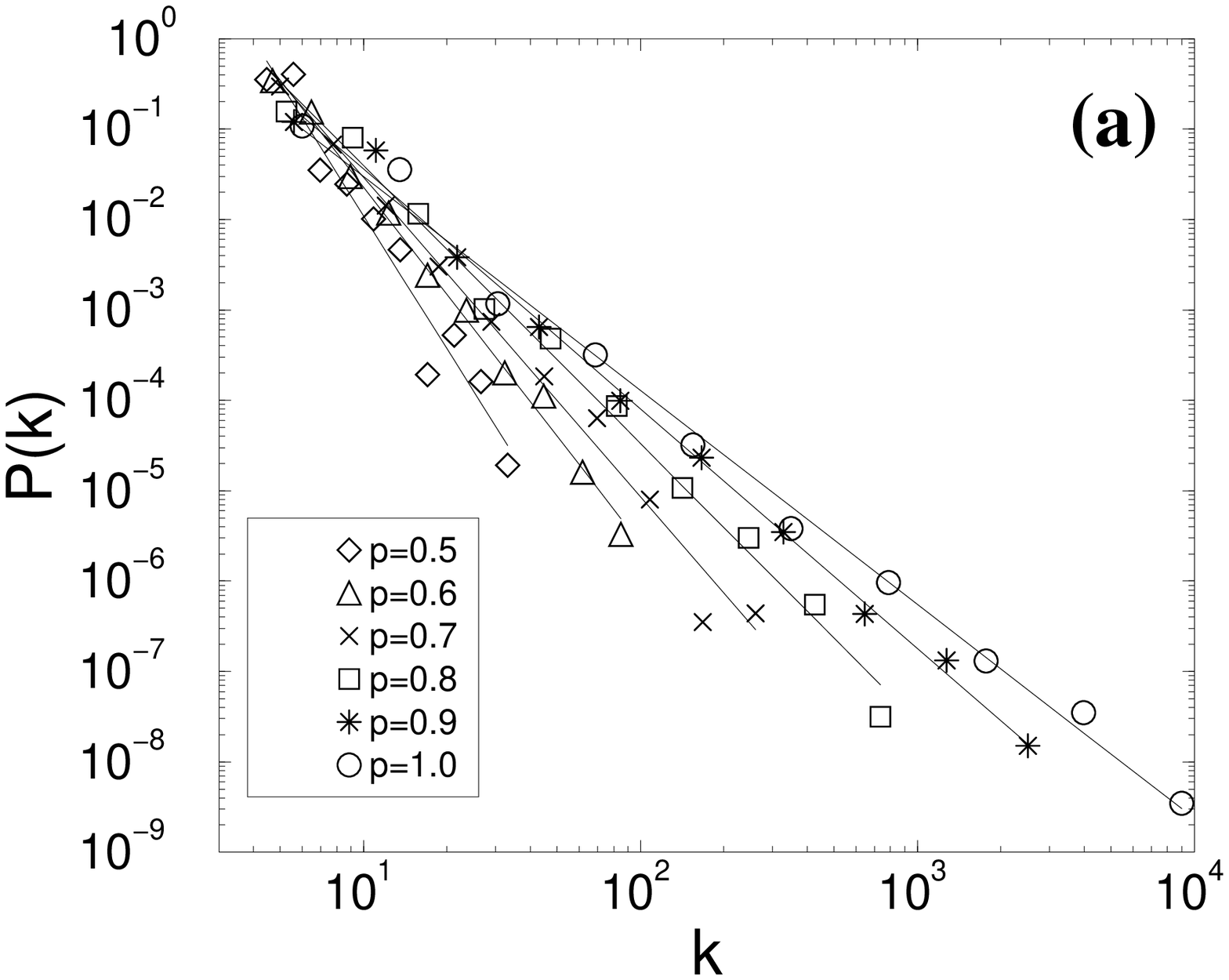}
     \end{minipage}
     \begin{minipage}[c]{0.49\textwidth}
	\centering
	\includegraphics[width=\textwidth]{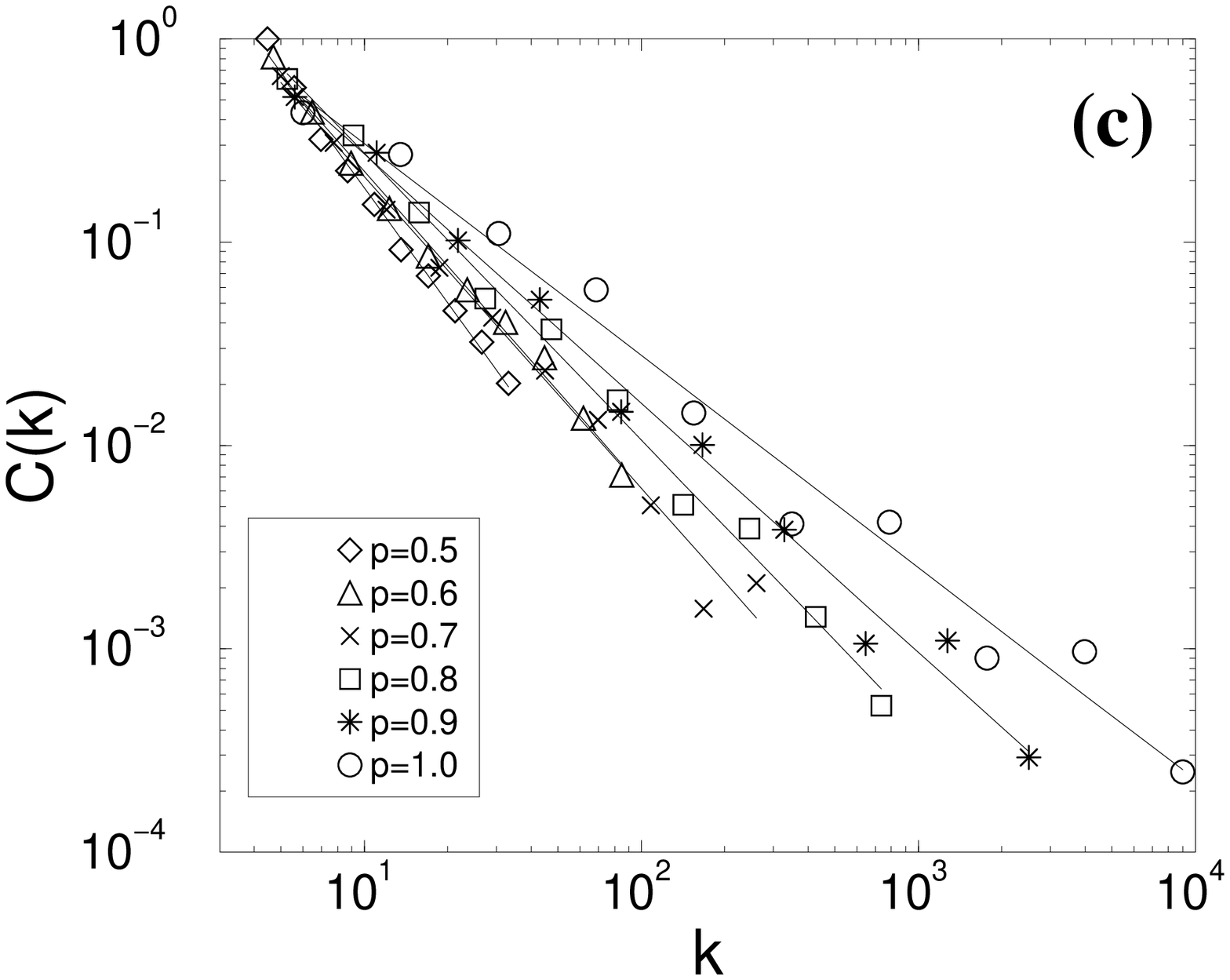}
     \end{minipage}
%
    \begin{minipage}[c]{0.49\textwidth}
	\centering
	\includegraphics[width=\textwidth]{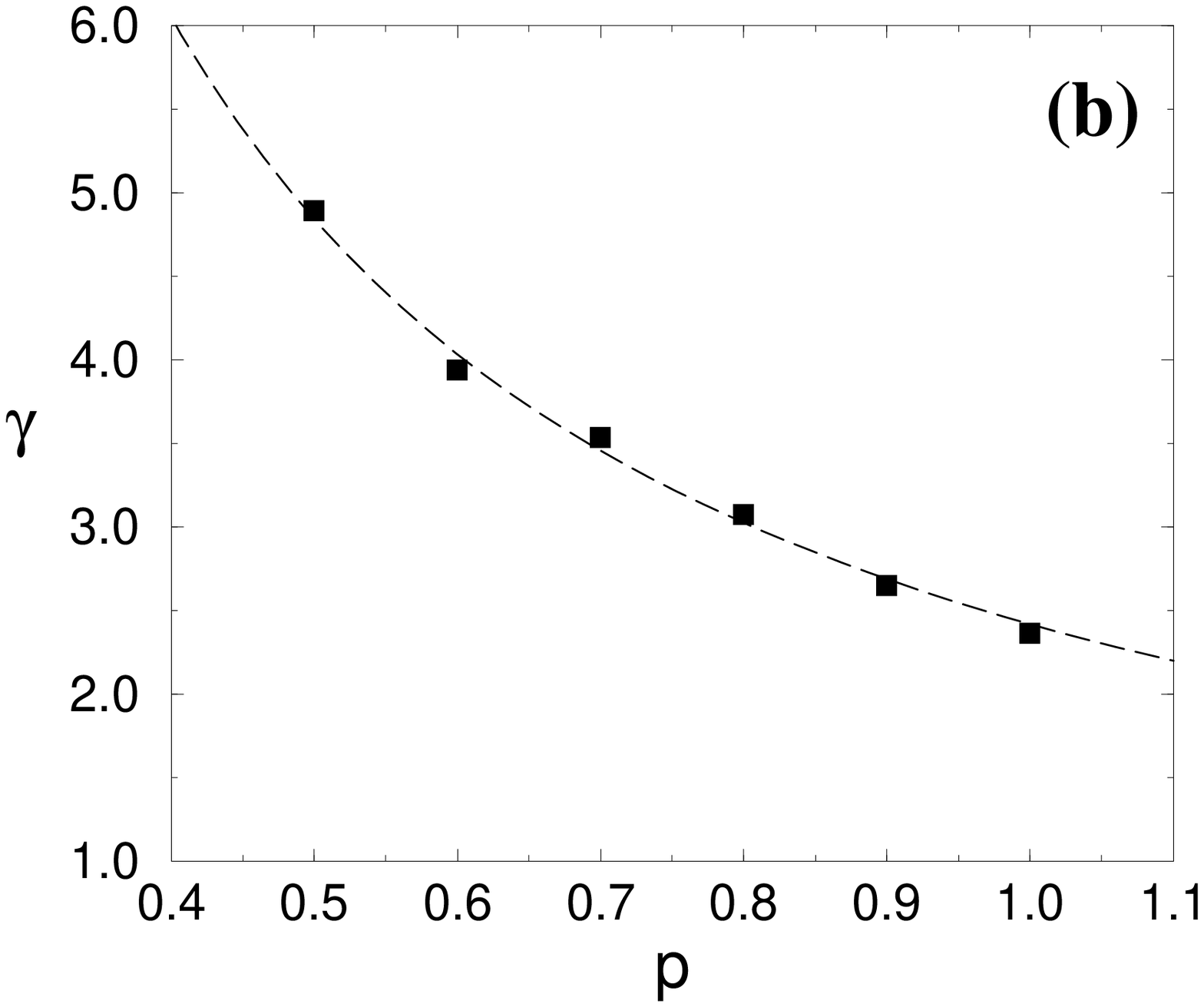}
     \end{minipage}
     \begin{minipage}[c]{0.49\textwidth}
	\centering
	\includegraphics[width=\textwidth]{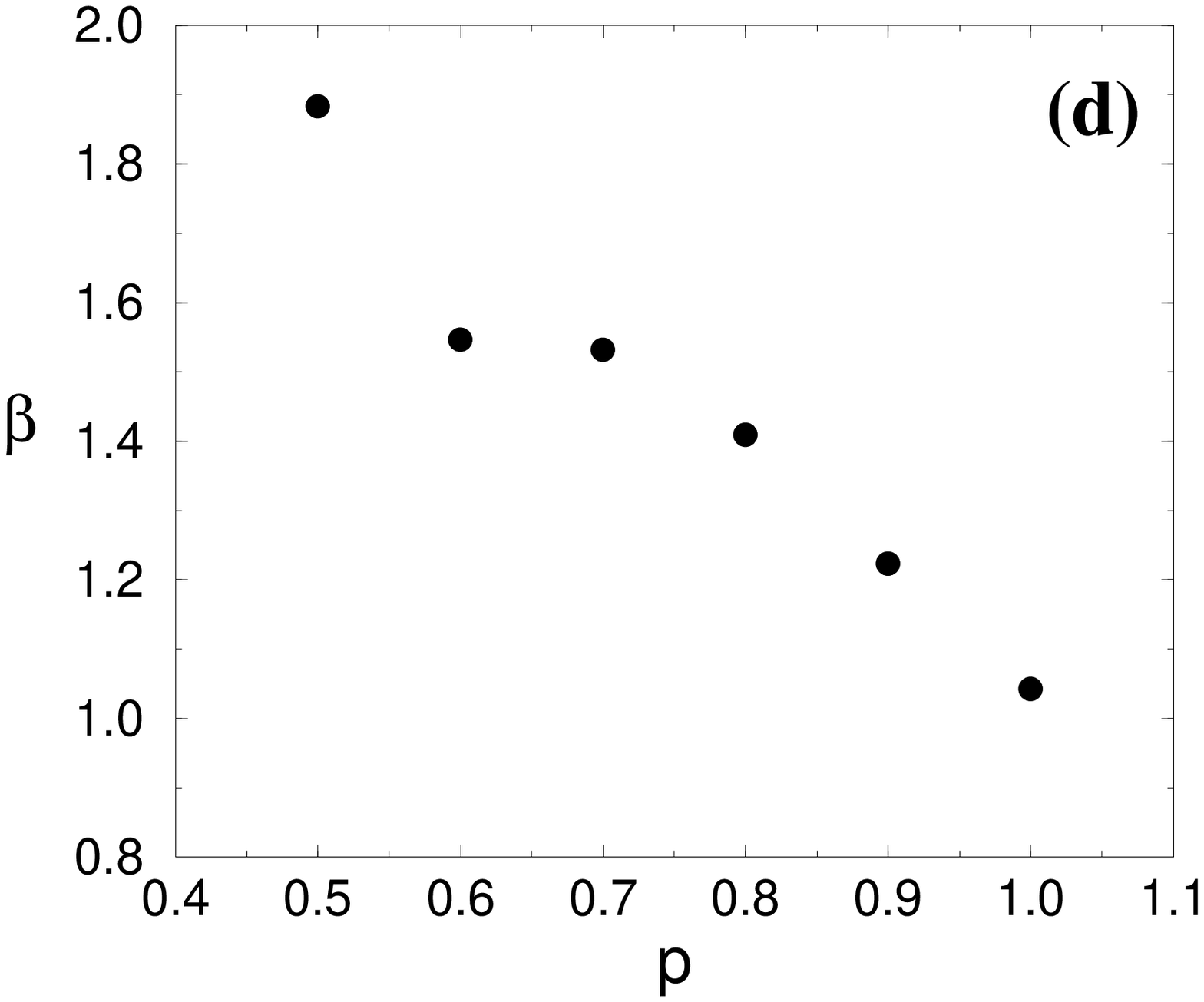}
     \end{minipage}
\end{minipage}

\caption{
	 The scaling properties of the stochastic model.
	{\bf (a)} The degree distribution for different $p$
		values, indicating that $P(k)$ follows a power law
		with a $p$ dependent slope.
	{\bf (b)} The dependence of the degree exponent $\gamma$ on $p$,
		determined by fitting power laws to the curves shown in {\bf (a)}.
			The exponent $\gamma$ appears to follow approximately
			$\gamma (p) \sim 1/p$ (dashed line).
	{\bf (c)} The $C(k)$ curve for different $p$ values,
		indicating that the hierarchical exponent $\beta$ depends on $p$.
	{\bf (d)} The dependence of  $\beta$ on the parameter $p$.
		  The simulations were performed for $N=5^7$(78,125) nodes.
	}
\label{fig:model}
\end{figure}

	As Fig.$\,$\ref{fig:model} shows, changing $p$
alters the slope of both $P(k)$ and $C(k)$ on a log-log plot.
	In general, we find that increasing $p$ decreases the exponents $\gamma$ and $\beta$
(Fig.$\,$\ref{fig:model}b,d).
The exponent $\beta=1$ is recovered for $p=1$, i.e. when all nodes of a module gain a link.
	While the number of links added to the network changes at each iteration,
for any $p\leq 1$ the average degree of the infinitely large network is finite.
	Indeed, the average degree follows
\begin{equation}\label{katlag}
	 \langle k \rangle_n= \frac{8}{5}\,\biggl(\frac{3}{2} + \frac{1-p^{n+1}}{1-p}\biggr),
\end{equation}
which is finite for any $p\leq 1$.

	Interestingly, the scaling of $C(k)$ is not a unique property
of the model discussed above.
	A version of the model, where we keep the fraction of selected nodes, $p$,
constant from iteration to iteration, also generates $p$ dependent $\beta$ and $\gamma$ exponents.
	Furthermore, recently several results indicate that the scaling
of $C(k)$ is an intrinsic feature of several existing growing networks models.
	Indeed, aiming to explain the potential origin of the scaling in $C(k)$
observed for the Internet, VSPV note that the fitness model
\cite{Ginestra_Euro,Ginestra_PRL} displays a $C(k)$ that appears to scale with $k$.
	While there is no  analytical evidence for $C(k)\sim k^{-\beta}$ yet, numerical results
\cite{vespin_internet_1,vespin_internet_CK_2} suggest that the
presence of fitness does generate a hierarchical network architecture.
	In contrast, in a recent model proposed by
Klemm and Eguiluz there is analytical evidence that the network obeys the scaling law (\ref{dgm_scale}) \cite{memory}.
	In their model in each time step a new node joins the network,
connecting to all \emph{active} nodes in the system.
	At the same time an active node is deactivated with probability $p\sim k^{-1}$.
	The insights offered by the hierarchical model
can help understand the origin of the observed $C(k)\sim k^{-1}$.
	By deactivating the less connected nodes
a central core emerges to which all subsequent nodes tend to link to.
	New nodes have a large $C$ and small $k$, thus they are rapidly deactivated, freezing into a large $C$ state.
	The older, more connected, surviving nodes are in contact
with a large number of nodes that have already disappeared from the active list, and they have small $C$
\cite{note_about_Vazquez_chain_hier_net}.

	Finally, Szab\'o, Alava and Kert\'esz have developed a rate equation method to
systematically calculate $C(k)$ for evolving networks models
\cite{structural_transition_Kertesz}.
	Applying the method to a model proposed by Holme and Kim \cite{Holme-Kim_model}
to enhance the degree of clustering coefficient $C$ seen in the scale-free model \cite{BA_model},
they have shown that the scaling of $C(k)$ depends on the parameter $p$, which governs the rate
at which new nodes connect to the neighbors of selected nodes, bypassing preferential attachment.
	As for $p=0$ the Holme-Kim model reduces to the scale-free model, Szab\'o, Alava and Kert\'esz
find that in this limit the scaling of $C(k)$ vanishes.
	These models indicate that several microscopic mechanisms could generate a hierarchical topology,
just as several models are able to create a scale-free network  \cite{reka_rev,dorog_rev}.

\section{Discussion and Outlook}

	The identified hierarchical architecture offers a new perspective on the topology of complex
networks.
	Indeed, the fact that many large networks are scale-free is now well established.
	It is also clear that most networks have a modular topology, quantified
by the high clustering coefficient they display.
	Such modules have been proposed to be a fundamental feature of biological systems
  \cite{hartwell,Erzso_SCI_Ecoli}, but have been discussed in the context of the WWW  \cite{giles,giles_web_clust},
and social networks as well  \cite{granovetter,WN_Sci}.
	The hierarchical topology offers a new avenue for bringing under a single roof these
two concepts, giving a precise and quantitative meaning for the network's modularity.
	It indicates that we should not think of modularity as the coexistence of
relatively independent groups of nodes.
	Instead, we have many small clusters, that are densely interconnected.
	These combine to form larger, but less cohesive groups,
which combine again to form even larger and even less interconnected clusters.
	This  self-similar nesting of different groups or modules into each other forces
 a  strict fine structure on real networks.

	Most interesting is, however, the fact that the hierarchical nature of these networks is well
captured by a simple quantity, the $C(k)$ curve, offering
us a relatively straightforward method to identify the presence of hierarchy in real networks.
	The law (\ref{dgm_scale}) indicates that the number and the size of the groups of different
cohesiveness is not random, but follow rather strict scaling laws.

	The presence of such a hierarchical architecture reinterprets the role of the hubs in
complex networks.
	Hubs, the highly connected nodes at the tail of the power law degree distribution, are
known to play a key role in keeping complex networks together, playing a crucial role from the robustness
of the network  \cite{attack,cohen_attack} to the spread of viruses in scale-free networks  \cite{vespin_virus_treshold}.
	Our measurements indicate that
the clustering coefficient characterizing the hubs decreases linearly with the degree.
	This implies that while the small nodes are part of highly cohesive, densely interlinked clusters,
the hubs are not, as their neighbors have a small chance of linking to each other.
	Therefore, the hubs play the important role of bridging the many small communities of clusters into a
single, integrated network.

\begin{acknowledgments}
	We benefited from useful discussions with J. Kert\'esz, Z. N. Oltvai and T.F Vicsek.
	We which to thank S. H. Yook and H. Jeong for providing us the language database.
	This research was supported by NSF, DOE and NIH.
\end{acknowledgments}


\newpage
\bibliography{hier_net_2col}

\end{document}